# Ethics for social robotics: A critical analysis

Júlia Pareto Boada[†]
Institut de Robòtica i Informàtica Industrial, CSIC-UPC
Facultat de Filosofia, Universitat de Barcelona (UB)
Barcelona, Spain
jpareto@iri.upc.edu

Begoña Román Maestre[†]
Facultat de Filosofia, Universitat de Barcelona (UB)
Barcelona, Spain
broman@ub.edu

Carme Torras[†]
Institut de Robòtica i Informàtica Industrial, CSIC-UPC
Barcelona, Spain
torras@iri.upc.edu

## ABSTRACT

Social robotics' development for the practice of care and European prospects to incorporate these AI-based systems in institutional healthcare contexts call for an urgent ethical reflection to (re)configurate our practical life according to human values and rights. Despite the growing attention to the ethical implications of social robotics, the current debate on one of its central branches, social assistive robotics (SAR), rests upon an impoverished ethical approach. This paper presents and examines some tendencies of this prevailing approach, which have been identified as a result of a critical literature review. Based on this analysis of a representative case of how ethical reflection is being led towards social robotics, some future research lines are outlined, which may help reframe and deepen in its ethical implications.

## 1 Introduction

As a technoscientific activity developing tools for specific fields of professional human activity, social robotics is a principal actor in the practical and conceptual (re)configuration of our life. Like all technoscientific advances, it modifies the margins of human action. Still, it does so in an unprecedented way by allowing us to "outsource" part of our agency to robots in human practices of a relational kind, such as care. Robots' capacity to interact with humans "interpersonally"[1] places social robotics as a promising technological contribution to European institutional care practices, mainly regarding healthcare[2][3]. Several European research initiatives and pilot projects[4][5] reveal significant prospects to incorporate social robots within professional contexts of (health)care, especially for assistance[3]. This scenario urges to engage in an ethical reflection that may contribute to normatively guiding social robotics' disruptive force already from the early and throughout all different stages of its growing development[6], with a view to the European ideal of a human-centered technology at the service of human rights and well-being[7]. Although the widespread proliferation of ethical discussion on social robotics shows an increasing awareness of this urge, there are some flaws in the predominant ethical approach; that is, in the perspective from which the ethical problems are identified, framed and addressed. This has to do with the newness of social robotics as a field of specific ethical attention[8] and a lack of a clear conceptual framework from which to engage in this normative-oriented kind of thinking.

In this paper, we present some main tendencies of the ongoing ethical reflection on social assistive robotics (SAR) disclosed through a critical literature review[9], and which indicate significant shortcomings of the prevailing ethical approach. By discussing these tendencies, we aim to lay some theoretical grounds for a comprehensive ethical approach to social robotics in general, and to point at some new ethics research directions regarding its development for (institutional) care practices.

## 2 The ethical debate on SAR: three tendencies

As identified through previous work[9], there are three significant tendencies of the current ethical reflection on SAR that should be redressed to enrich the debate on the implications of this technoscientific field of activity. In the following subsections, we set them out and briefly argue why they entail an impoverished ethical approach.

### 2.1 An individual-centered perspective

The ongoing ethical reflection on SAR is predominantly led from an individual-centered perspective, which focuses on the implications that robots may have for the well-being of humans with whom they interact. Much limited to the sphere of human-robot interaction (HRI), which is furthermore inadequately comprehended in dyadic terms[10][11], this ethical approach comes along with less attention to SAR implications from both the perspective of the specific (care) practice in which AI systems are introduced and the sociopolitical perspective of justice. This tendency means an important deficiency for a proper ethical approach, since it overlooks the constitutive interrelation between individual Well-being, Care and Justice as the main spheres of human activity with ethical importance regarding SAR. Indeed, an excessively restricted ethical focus at the individual level of human life and framed on the dyadic interaction between humans and robots implies overlooking the role of sociopolitical structure in the configuration of care practices, and thus the influence that these both have regarding individual Well-being. Thereby, this tendency unjustifiably falls in the "neglect of the political" underlying the mainstream philosophy of technology and ethics of technology, which some authors have already condemned and contributed to redressing[12].

## 2.2  A narrowed understanding of ethical concepts

The individual-centered perspective seems to be tightly correlated to a second significant tendency of the ethical debate: a narrowed understanding of certain core ethical concepts around which SAR's problems stand, such as freedom and its related concept of autonomy, as well as responsibility. (Human) freedom is currently much restricted to what is philosophically known as "negative liberty". Philosophical accounts of freedom[13][14] and autonomy[15] have, though, a richer scope of meaning, which enables a transversal gaze to SAR implications at the *micro*, *meso* and *macro* level of human life. The same happens with responsibility, which is currently understood in the traditional sense of liability for harm; a harm which is moreover linked to the robot's behavior. Philosophical approaches to responsibility offer a broader and more "substantial" understanding of the notion[16], according to which the ethical approach would be also framed in terms of accountability for technological development[17], thereby bringing to the fore discussion on the teleology and interests to which it is linked (perspective of justice). Therefore, the current restricted understanding of these notions comes along with an impoverishment of the ethical reflection.

## 2.3  A lack of discussion on SAR teleology

A third tendency of the current ethical approach is a general lack of explicit discussion on SAR teleology, that is, the "ends" at which it aims to serve, the "what for" of its development. Conceptual assumptions on "care" and "assistance", as well as other correlated notions that underlie the field's development (human well-being, human capabilities, autonomy) are usually not openly examined and revised, although some authors have already evinced the need of and engaged in such analysis[18][19]. This means an important deficit in the predominant ethical reflection on SAR. The reason is to be found in the instrumental nature of technology, that is, in the fact that technology ultimately has an end that is external to itself, in the sense that its goal is to serve the purposes of the activities for which it is conceived as a means of support. This demands to conduct ethical reflection on SAR primarily as an exercise of applied ethics[20], understood as critical hermeneutics of human activity[21]. That is, the ethical implications of SAR must primarily be thought in the light of the specific practice for which technology is conceived, within the framework of its defining goals and values. For instance, the ethical issues of HRI must be examined in the light of the particular practice within which this interaction takes place: the conflicts of value coming into play in HRI will not be the same whether this interaction happens in the framework of a service activity (reception in a hotel) or of a care practice, such as when the interaction is a means to provide assistance in cognitive rehabilitation or company in front of solitude. Since SAR aims at contributing to practices of care, the scarce discussion on the constellation of SAR teleological-related meanings, altogether with the predominant ethical individual-centered perspective (instead of an approach primarily focused on the implications of SAR from the care-practice perspective) accounts for an impoverished ethical reflection.

## 3  Social robotics for care: towards a comprehensive ethical approach

Social robotics' development for care practices requires a comprehensive ethical approach that identifies and analyses the implications of this technoscientific field of activity at different levels of human life. As contended, this means, primarily, an exercise of applied ethics –in which reflection is contextualized according to social robotics' practical field of application–, coming along with a critical theory perspective –through which not only means, but also ends, are included in the discussion (which type of practices, societies and lives are to be reconfigured through social robotics' development?)–. Thus, an exhaustive ethical approach also demands to address social robotics' development from the macro perspective of justice. All this requires an ethical gaze that takes into account conceptual advances of other subdisciplines of philosophy besides ethics, such as the refutations of the technology's value-neutrality thesis coming from the philosophy of technology[22][23][24], as well as the disclosure of the political dimension of technology set forth by political philosophy[25].

To achieve such a broader and deeper normative thinking on the ethical implications of social robotics for (institutional) care practices, we outline two main research lines to be next developed, considering the analyzed state of the art.

### 3.1  Approaching social robotics from a philosophical account of freedom and autonomy

Unfolding the philosophical concept of freedom and autonomy and ethically (re)examining social robotics in the light of these will broaden the ethical implications of the latter, by means of bringing to the fore the interdependence between freedom and the sociopolitical structuring of human life, as well as the political-structural dimension of human-technology relations. Issues concerning domination, manipulation or increased vulnerability raised by social robotics will appear not only regarding the interpersonal level of human life (linked to HRI in traditional terms) but also the structural one. This line of research could (partially) contribute to redressing the current individual-centered perspective.

### 3.2  Examining social robotics from the idea of "care"

Reflecting upon the notion of care as a practice (mainly drawing from J. Tronto's work [26]) and identifying the ethical considerations regarding social robotics' development that emerge from this point of view will contribute to a proper exercise of (applied) ethics, by suitably redressing the primary focus of ethical attention to the sphere of the practice and enabling to filter reflection in the light of it. Moreover, analyzing the idea of care will help clarify and discuss the (shared) grounds of many ethical issues associated to HRI in assistive contexts (such as deception, dignity, emotional attachment, unauthentic intersubjectivity, objectification and recognition)[27], as well as the technoscientific[28] and institutional corresponding narratives on innovation and implementation of social robots for care[29]. By disclosing what is potentially at stake with the introduction of social robots in care practices, this analysis will help delineate the sort of ethical considerations that should be reflected upon when developing this technology for institutional practices of care.

## ACKNOWLEDGMENTS

This work has been partially supported by the Spanish Ministry of Science and Innovation under a FPI scholarship for predoctoral contracts for the training of doctors (PRE2018-084286), and by the European Union Horizon 2020 Programme under grant agreement no. 741930 (CLOTHILDE).